\documentstyle[aps,epsf,floats]{revtex}
\def\ge{\hbox{$\;\raise.4ex\hbox{$>$}\kern-.75em\lower.7ex\hbox{$\sim$}
                    \;$}}
\def\grsize{3.5cm}
\begin{document}

\title{Random matrix study of the phase structure of QCD with two colors}

\author{Beno\^\i t Vanderheyden$^{1,2}$ and A. D. Jackson$^1$}

\address{
$^1$The Niels Bohr Institute, Blegdamsvej 17, DK-2100 Copenhagen \O, Denmark.\\
$^2$Institut d'Electricit\'e Montefiore, Campus du Sart-Tilman B-28, B-4000
Li\`ege, Belgium.}

\date{\today}

\maketitle

\begin{abstract}
  
  We apply a random matrix model to the study of the phase diagram of QCD
  with two colors, two flavors, and a small quark mass.  Although the effects
  of temperature are only included schematically, this model reproduces most
  of the ground state predictions of chiral perturbation theory and also
  gives a qualitative picture of the phase diagram at all temperatures. It
  leads, however, to an unphysical behavior of the chiral order parameter and
  the baryon density in vacuum and does not support diquark condensation at
  arbitrarily high densities.  A better treatment of temperature dependence
  leads to correct vacuum and small temperature properties. We compare our
  results at both high and low densities with the results of microscopic
  calculations using the Nambu-Jona-Lasinio model and discuss the effects of
  large momentum scales on the variations of condensation fields with
  chemical potential.

\end{abstract}

\section{Introduction \label{s:intro}}

A number of early and recent model
calculations~\cite{early,RapSch98,AlfRaj98,review} indicate that
quark-quark correlations may play an important role in QCD at finite
density. By inducing pairing gaps $\Delta \sim 100$ MeV, such
correlations may prove relevant for the physics of neutron and compact
stars as well as that of heavy-ion
collisions~\cite{RapSch98,AlfRaj98,review,neutron1,neutron2,CarRed00,AlfBr00}.
Due to the limitations of standard Monte Carlo techniques when applied
to systems with finite baryon density, the study of these effects
through lattice simulations is very challenging.  A non-zero chemical
potential leads to a complex determinant of the Euclidean Dirac
operator and results in massive cancellations among configurations.
This difficult problem stands in the way of a qualitative
understanding of the critical physics involved in lattice
simulations~\cite{BieConf98}.

One response to this situation is to consider QCD-like theories with
additional antiunitary symmetries that guarantee the reality of the
fermion determinant and therefore render such theories more amenable to
lattice formulations.  Examples of such theories include QCD with an
arbitrary number of colors, $N_c$, and adjoint quarks as well as QCD
with two colors and fundamental quarks.  Knowledge of the critical
physics in these cases may offer clues regarding the critical
physics of the more difficult problem of QCD with three colors.

Here, we will consider QCD with two colors.  In this case, quark and
antiquark states transform similarly under global color rotations.
They can be combined into spinors with an extended flavor symmetry
${\rm SU}(2 N_f)$ for which $\left<qq\right>$ baryons and $\left<\bar
  q q\right>$ mesons belong to the same
multiplets~\cite{DiaPet92,SmiVer95}. In particular, the lightest
baryons and the pions have a common mass, $m_\pi$.  This spectrum
determines the properties of the ground state for small chemical
potential $\mu >0$. General arguments~\cite{HalJac98} indicate that a
transition from the vacuum to a state with finite baryon density
should take place at a critical chemical potential, $\mu_c$, which is
the lowest ratio of energy to baryon number that can be realized by
an excited state of the system.  Here, this state is populated by
light $\left<qq\right>$ baryons, and the transition is thus expected
to take place at $\mu_c = m_\pi/2$.  The $(T,\mu)$ phase diagram of
QCD with two colors has been studied previously by Dagotto et al.\ 
using a mean-field model of the lattice action~\cite{DagKar86}.  Their
results confirm the transition at $\mu_c$.  More recently, the
smallness of $\mu_c \sim m_\pi/2$ has been exploited in studies of the zero
temperature phase transition using chiral perturbation theory extended
to the flavor symmetry ${\rm SU}(2
N_f)$~\cite{KogSteTou99,KogSteTou00}.  Many of these model
calculations have been verified by recent lattice
simulations~\cite{HanMon00,SplSon00,AloGal00,MurNak00,BitLom00,Lom99}.

We have recently constructed a random matrix model of QCD with three
colors which permits thermodynamic competition between chiral and
diquark condensation~\cite{VanJac99}.  The extension of this model to
QCD with two colors is straightforward and constitutes a useful
consistency check. The random matrix model respects the global flavor
and color symmetries of the QCD interactions and eliminates much of
their remaining detailed structure.  For two colors, this model offers a
qualitatively correct picture of the phase diagram but, at first
sight, appears to fail in two respects.  First, in it simplest 
implementation in which only the two lowest Matsubara frequencies are 
retained, we find an unphysical behavior of the baryon density and the chiral
condensate in vacuum.  Second, the random matrix interactions
saturate for large $\mu$ and do not support diquark condensation at
arbitrarily high baryon density.

Our primary goal in this study is to examine these difficulties and
their relative importance in establishing a reliable picture of the
phase diagram.  The first difficulty (i.e. the ground state behavior
of the baryon density and the chiral condensate) can be resolved by
including all quark Matsubara frequencies, which is equivalent to
the introduction of proper Fermi occupation factors.  For small $T$, the
full frequency sum also modifies the temperature dependence of the onset
chemical potential from a quadratic law with $\mu_c(T)-\mu_c(0) \sim
T^2$ to essential singular behavior with $\mu_c(T)-\mu_c(0) \sim
\exp (-\Sigma/T)$, where $\Sigma$ is the vacuum chiral condensate. The
remaining phase diagram is qualitatively unchanged. The Matsubara sum
is equivalent to a monotonic mapping of the phase diagram of the
original random matrix model, a mapping which therefore preserves the
topology of the phase structure. 

The second difficulty (i.e. the saturation of the interactions at
large $\mu$) results from the lack of a true Fermi sea of states in
random matrix models.  We consider the effects of a proper Fermi sea
by studying a microscopic model with an interaction similar to that of
the Nambu-Jona-Lasinio (NJL) model.  This model naturally
exhibits diquark condensation at all densities. It also yields an
additional power law correction to the essential singular dependence
of $\mu_c(T)$.  It will be shown that other deviations from the prediction 
of the random matrix model arise only in regions where the
condensates are weak and thus sensitive to the detailed form of the
interaction.  Also related to non-universal features of the
interactions is the observation that deviations from the results of
chiral perturbation theory for $\mu$ on the order of $\mu_c$ are 
determined by the form of the interaction on momentum scales substantially 
larger than $\mu_c$.  This is true for both random matrix and NJL models.  
The sign of these deviations depends on model details and cannot in general be
established from first principles.

The remainder of this paper is organized as follows.  We review the
random matrix model of Refs.~\cite{VanJac99}, extend it to the case
of two colors, and discuss its low temperature properties in
Sec.~\ref{s:model}. We analyze the sum over all Matsubara frequencies in
Sec.~\ref{s:all} and study the NJL model in Sec.~\ref{s:fermi}. We discuss
the phase diagrams and the condensation fields for each case and present our
conclusions in Sec.~\ref{s:conclusions}.

\section{The random matrix model}
\label{s:model}

We first consider the random matrix model of Refs.~\cite{VanJac99} in
a form appropriate for two colors and two flavors. In this model, the
single-quark Lagrangian is represented by a matrix whose block
structure reflects the global chiral and color symmetries of QCD. The
matrix elements describe the degrees of freedom associated with the
background of gluon fields. Aside from the constraints imposed by the
prescribed block structure, all matrix elements are independent and
chosen at random.  An exact solution can be obtained if their distribution
is Gaussian. Then, an integration over this distribution produces a 
partition function of the form
\begin{eqnarray}
Z(\mu,T) & = & \int d\sigma\,\,d\Delta \,\, 
\exp\left[- N \Omega(\sigma,\Delta)\right],
\label{part_function}
\end{eqnarray}
where $T$ is the temperature and $\mu$ is the chemical potential.
Here, $N$ is proportional to the matrix size and represents the number
of low-lying degrees of freedom involved in the phase transition. This
number scales with the volume of the physical system, and the
thermodynamic limit therefore corresponds to $N \to \infty$.  The
integrals in Eq.~(\ref{part_function}) are performed over 
auxiliary chiral and diquark fields, $\sigma$ and $\Delta$.  We
consider here the familiar chiral channel, $\sigma \sim \left<\bar q
  q\right>$, and the diquark channel, $\Delta \sim \left<q\, C\gamma_5
  \, \tau_2 \,\tau_2^{F} \, q\right>$, where $C$ is the charge
conjugation matrix ($C \gamma_\mu C = \gamma_\mu^T$) and where
$\tau_2$ and $\tau^{\rm F}_2$ are the antisymmetric ${\rm SU}(2)$
color and flavor generators~\cite{KogSteTou00}. (For a related random
matrix model which takes into account additional effects from
instanton--anti-instanton molecules, see \cite{PepSch00}.)

For single-gluon exchange, the thermodynamic potential $\Omega$
assumes the form
\begin{eqnarray}
\Omega(\sigma,\Delta) = A (\sigma^2+\Delta^2) - {\rm Tr} \log
S(\sigma,\Delta) + \Omega_{\rm reg},
\label{Omega}
\end{eqnarray}
where $A$ characterizes the coupling strengths of the interactions in
the chiral and diquark channels.  (These  strengths are equal
for single-gluon exchange.)  The quantity $S(\sigma,\Delta)$ is the
propagator for a single quark in the background of the condensation
fields, $\sigma$ and $\Delta$.  The trace is performed over
the quark Matsubara frequencies $\omega_n = i \mu + (2 n + 1) \pi T$,
where $n$ is an integer running from $-\infty$ to $\infty$.

The random matrix model describes only the thermodynamic contribution
from low-energy modes. The effects of the remaining high-energy modes can, 
in principle, lead to a ``regular'' term, $\Omega_{\rm reg}$, which does
not affect the critical physics.  This term is independent of $\sigma$
and $\Delta$ and is an analytic function of $T$ and $\mu$.  Although
the form of $\Omega_{\rm reg}$ cannot be determined by random matrix
arguments, we retain it as a reminder of the presence of
high-energy modes.  We will later exploit this degree of freedom to
adjust the small temperature properties of the model.

In its simplest implementation, the model retains only the two lowest 
Matsubara frequencies $n = \pm 1$. The trace in Eq.~(\ref{Omega}) is then 
reduced to
\begin{eqnarray}
{\rm Tr} \log S(\sigma,\Delta) & = & \sum_\pm \log(E^2_\pm(\sigma,\Delta)+\pi^2
T^2), 
\label{Tr_log_one}
\end{eqnarray}
where $E_\pm(\sigma,\Delta)$ represent the quark ($-$) and antiquark ($+$)
excitations energies in the background of the condensation fields,
\begin{eqnarray}
E_\pm(\sigma,\Delta) & \equiv & \left((\sigma+m \pm \mu)^2+
\Delta^2\right)^{1/2}.
\label{Epm:one}
\end{eqnarray}
The simplest model thus consists of Eqs.~(\ref{Omega}),
(\ref{Tr_log_one}) and (\ref{Epm:one}).  We note that a similar form
of the potential $\Omega (\sigma,\Delta)$ is also obtained for a
chiral random matrix theory with a Dyson index $\beta =
1$~\cite{chiRMM}, in which one retains only the chiral and diquark
channels.  In contrast to the present model, this theory makes
explicit use of the pseudoreality of QCD with two colors by choosing
real matrix elements. However, it does not refer explicitly to color
and spin quantum numbers. As a result, the diquark condensate is
different from that in the present model and develops instead in the
$\left<q^T \gamma_5 q\right>$ channel. Since the two models possess
interactions with similar symmetries among their respective
condensates, they naturally lead to similar potentials $\Omega
(\sigma,\Delta)$.

Returning to our model, we first consider the chiral limit, $m=0$.
Then, the condensation fields in vacuum ($\mu = T = 0$) appear only in
the combination $\sigma^2+\Delta^2$, and the system obeys the extended
flavor symmetry ${\rm SU}(4)$. The spontaneous formation of a chiral
condensate, $\sigma \neq 0$ and $\Delta = 0$, further breaks the
symmetry down to ${\rm Sp}(4)$.  A pure chiral solution,
$(\sigma,\Delta) = (\Sigma,0)$, can be transformed into a pure diquark
solution, $(\sigma, \Delta) = (0,\Sigma)$, by an ${\rm SU}(4)$
rotation.  These two solutions are thus thermodynamically
indistinguishable, and they describe a single phase with $\sigma^2+\Delta^2
= \Sigma^2 = 2/A$. This phase persists up to a critical temperature
$T_{c} = \Sigma/\pi$ above which all condensation fields vanish and
the ${\rm SU}(4)$ symmetry is restored via a second-order transition.
(This is a mean-field result; renormalization group arguments show
that for two or more flavors, fluctuations may actually drive the
transition first-order~\cite{Wir99}.)  For $T=0$, a positive chemical
potential, $\mu > 0$, breaks the ${\rm SU}(4)$ symmetry down to ${\rm
  SU}_{\rm L}(2) \times {\rm SU}_{\rm R}(2)\times {\rm U}(1)_{\rm B}$; a
quark mass $m\neq 0$ breaks it further to ${\rm SU}_{\rm V}(2)\times
{\rm U}(1)_{\rm B}$~\cite{KogSteTou99,KogSteTou00}.  In the following, we
consider light quarks by choosing $m \ll \Sigma$, where
$\Sigma=\sqrt{2/A}$ is the vacuum chiral condensate for $m=0$.  Note
that for $m \neq 0$, the chiral field is no longer a true order
parameter, and the second-order phase transition at $\mu=0$ becomes 
a cross-over.

The diquark field, $\Delta$, remains a true order parameter even when
$m \ne 0$.  The system favors the formation of diquark pairs in
regions where $\partial^2 \Omega/\partial \Delta^2(\sigma,0) < 0$, and
a pure chiral solution can be a saddle-point of the potential at
best. There is therefore a second-order phase transition along the
line $\partial^2 \Omega/\partial \Delta^2(\sigma(\mu,T),0) = 0$,
where $\sigma(\mu,T)$ obeys the chiral gap equation $\partial \Omega
(\sigma,0)/\partial \sigma =0$.  This line is given by the relation
\begin{eqnarray}
\mu^2+\pi^2 T^2 & = & {\Sigma^2 \mu^2 \over \mu^2-m^2} - {\Sigma^4 m^2 \over 
4 (\mu^2 - m^2)^2}
\label{insline}
\end{eqnarray}
and delimits a region in the $(T,\mu)$ plane as illustrated in Fig.~1~$(a)$.  
(We will discuss Figs.~1~$(b)$ and $(c)$ below.)  Its intercept with the 
$T=0$ axis determines the onset chemical potential 
\begin{eqnarray}
\mu_c^2 = {m \Sigma \over 2} + {\cal O}(m^2),
\label{mu_c}
\end{eqnarray}
which becomes $\mu_c(T) \sim \mu_c (1+ \pi^2 T^2/(4 \Sigma^2))$ for small
$T$.  According to the arguments in the introduction, the onset chemical
potential of Eq.~(\ref{mu_c}) also obeys $\mu_c=m_\pi/2$. This leads us to
identify the pion mass as $m_\pi \simeq m^{1/2} \Sigma^{1/2} + {\cal O}(m)$.
 
This result is not trivial and may be regarded as the random matrix
analogue of the current-algebra relation $m_\pi \sim m^{1/2}$. To
understand the origin of this relation in the present model, consider
the random matrix vacuum.  For simplicity, assume one quark flavor and
ignore diquark condensation by setting $\Delta = 0$. Consider first $m
= 0$. In this case, we know that the spontaneous breaking of chiral
symmetry guarantees the existence of a massless Goldstone
mode~\cite{CarJac99} independent of the number of colors~\cite{rem}.
Consider next the case of a finite but small mass $m$, for which
chiral symmetry is explicitly broken. The previous Goldstone mode now
describes the lowest-lying fluctuations around the global minimum of
$\Omega(\sigma,0)$, which is given by $\sigma_0 = \Sigma-m/2+{\cal
  O}(m^2/\Sigma)$.  From Eqs.~(\ref{Omega}), (\ref{Tr_log_one}), and
(\ref{Epm:one}) with $\Delta = 0$, we find that the energy of fluctuations
around this solution is given as
\begin{eqnarray}
\Omega_{\rm fluct} & \simeq & 
{2\over \Sigma^2} \left((\sigma_0+ \delta \sigma)^2 + \delta \pi^2\right)
  - 2
  \log\left[\left(\sigma_0 + m+\delta\sigma \right)^2+\delta\pi^2\right],
\label{Omega_fluct1}
\end{eqnarray}
where the pion mode $\delta\pi$ describes oscillations in the
direction with $\left<\bar q \gamma_5 q\right>\neq 0$ and $\delta
\sigma$ describes those in the $\left<\bar q q\right>$ direction. Thus
the mass, $m$, acts as an external field with an almost linear
coupling to the order parameter; for small $m$, the symmetry breaking
term goes as $\delta\Omega_{\rm fluct} \propto - m\, \sigma_0 $. This
coupling causes the curvature of the free-energy in the soft
direction to be proportional to $m$.  Expanding to second order in
$\delta\sigma$ and $\delta \pi$ and working to lowest order in
$m/\Sigma$, we find
\begin{eqnarray}
\Omega_{\rm fluct}& \simeq & \left. \Omega\right|_{\delta\sigma=\delta\pi=0}
+ {4\over \Sigma^2}\, \delta \sigma^2 +
{2 m\over \Sigma^3}\, \delta \pi^2 + \cdots
\label{Omega_fluct}
\end{eqnarray}
This allows us to identify the mass ratio $m_\pi^2/m_\sigma^2$ as
$\sim m/\Sigma$ and deduce $m_\pi^2 \sim m$, which confirms the result
of Eq.~(\ref{mu_c}).  The generalization of this argument to include
fluctuations in the $\left<qq\right>$ channel is straightforward and
leads to results similar to those of chiral perturbation theory for 
the constant terms~\cite{KogSteTou00}.  (Random matrix theory cannot 
reproduce terms with spatial derivatives.) 

We now return to the diquark condensed phase. The two gap equations,
${\partial \Omega / \partial \sigma}  =  0$ and
${\partial \Omega / \partial \Delta}  = 0$,
possess a single solution with both $\sigma \neq 0$ and $\Delta \neq 0$,
\begin{eqnarray}
\sigma & = & -m + {\Sigma^2 \over 2} {m\over \mu^2-m^2}, 
\label{mix_fields1} \\
\sigma^2+\Delta^2 & = & \Sigma^2 + m^2 -\mu^2 - \pi^2 T^2,
\label{mix_fields}
\end{eqnarray}
where $\Sigma \equiv (2/A)^{1/2}$.  Expanding to the lowest orders in 
$m/\Sigma$, the fields near the condensation edge $\mu \approx \mu_c$ obey
\begin{eqnarray}
\sigma & \simeq & \Sigma\,\, {\mu_c^2 \over \mu^2} - m +
{\cal O}\left({m^2 \over \Sigma}\right), \\
\sigma^2 + \Delta^2 & =  & \Sigma^2  - \pi^2 T^2  + {\cal O}(m \Sigma). 
\label{mix_fields_exp}
\end{eqnarray}
For $T=0$ and to leading order in $m/\Sigma$, these results coincide
with chiral perturbation theory~\cite{KogSteTou00}. For $\mu > \mu_c$,
the chiral field decreases like $1/\mu^2$, and this saddle-point
solution rotates into a pure diquark solution keeping
$\sigma^2+\Delta^2$ nearly constant.  Corrections of order $m\Sigma$
in Eq.~(\ref{mix_fields_exp}) depend on the explicit functional form
of the potential of Eq.~(\ref{Omega}) (i.e.\ here, a logarithm
involving certain combinations of $\sigma$ and $\Delta$) and are thus
not universal.  We will see below examples of other functional
dependences which give different corrections in $m\Sigma$.  If we now
turn to the non-condensed phase ($\mu < \mu_c$ and $\Delta = 0$), we
find
\begin{eqnarray}
\sigma \simeq \Sigma - {m \over 2} + {\mu^2 - \pi^2 T^2 \over 2 \Sigma} + 
{\cal O}\left( {m^2 \over \Sigma} \right),
\label{chiral:one}
\end{eqnarray}
where terms of order $m$ and higher are again not universal.

Although the results to lowest order in $m/\Sigma$ are consistent with
our expectations, higher-order terms give rise to certain unphysical
properties. For $T=0$ and $\mu < \mu_c$, we expect that, in the
absence of confinement, the lowest-lying excitations carrying a net
baryon number are single quarks with an energy $\sim \Sigma$.  Below
the condensation edge, we have $\mu < \mu_c \ll \Sigma$, and the
excitation energy $\sim \Sigma$ is well above the Fermi level $\mu$.
The ground state should thus have a chiral field equal to its value at
$\mu = 0$ and a baryon density of zero.  We find instead a chiral
field $\sigma \simeq \Sigma -m/2 +\mu^2/(2 \Sigma)$, which increases
quadratically with $\mu$.  The behavior of the baryon density, $n_{\rm
  B} \equiv -\partial \Omega/\partial \mu$, is also incorrect.  From
the solutions above, we find that the potential in each phase obeys
\begin{eqnarray}
\Omega & \simeq & \Omega_{\rm vac} - { 4 m\over \Sigma} +
2\,\, {\mu^2 - \pi^2 T^2 \over \Sigma^2} + \Omega_{\rm reg} +
{\cal O} \left({m^2\over \Sigma^2}\right) \quad (\mu < \mu_c), \\
\Omega & \simeq & \Omega_{\rm vac} - {m^2\over \mu^2} -
2\,\, {\mu^2+ \pi^2 T^2 \over \Sigma^2} + \Omega_{\rm reg} +
{\cal O}\left({m^2\over \Sigma^2}\right)
\quad (\mu \ge \mu_c),
\label{Omegas:one}
\end{eqnarray}
where the vacuum energy is $\Omega_{\rm vac}= 2 - 2 \log\Sigma^2$. These
relations lead to negative baryon density for $\mu < \mu_c$! 

This behavior is a result of having truncated the Matsubara sum in
Eq.~(\ref{Omega}) to the two lowest frequencies.  This approximation 
does not affect the critical physics but does modify bulk properties.  It 
is therefore appropriate to correct the shortcomings of such an 
approximation naturally with an appropriate adjustment of the regular term, 
$\Omega_{\rm reg}$, in the thermodynamic potential of Eq.~(\ref{Omega}).  
The requirement that the baryon density should vanish for $\mu < \mu_c$ is 
readily met by the choice $\Omega_{\rm reg} = - 2
\mu^2/\Sigma^2$.  For $T=0$, we then obtain a baryon density of 
\begin{eqnarray}
n_{\rm B} =  -{\partial \Omega \over \partial \mu} & = & \left\{
\begin{array}{cc}
0 + {\cal O}\left({m \mu_c\over \Sigma^3}\right) & (\mu < \mu_c) \\
{ 8 \mu / \Sigma^2} \left(1-{\mu_c^4 /\mu^4} \right)
+ {\cal O}\left({m \mu_c\over \Sigma^3}\right)
& (\mu > \mu_c)\, , \\
\end{array} \right. 
\label{nb_one}
\end{eqnarray}
which has the same $\mu$-dependence as that found in chiral
perturbation theory~\cite{KogSteTou00}. For $\mu < \mu_c$, the terms
of order ${\cal O}({m \mu_c/ \Sigma^3})$ are positive, hence $n_{\rm B} >
0$ for all $\mu$.

To summarize, the random matrix model reproduces the results of chiral
perturbation theory to leading order in $m/\Sigma$ for the
condensation fields and for the baryon density, provided only that we
exploit the freedom of including a component $\Omega_{\rm reg}$ in the
thermodynamic potential.  This addition cannot, however, affect the
condensation fields, and $\sigma$ still grows weakly with $\mu$ below
the condensation edge. The variation of the condensation fields near
the edge is shown in Fig.~2~$(a)$. (Figs.~2~$(b)$ and $(c)$ will be
discussed below.) It is, in fact, difficult to discern the dependence of
$\sigma$ on $\mu$ for $\mu < \mu_c$ as the quadratic term $\sim
\mu^2/\Sigma$ is of order $m/\Sigma$ relative to the leading term, which 
is of order $\sim \Sigma$.  Figure~3~$(a)$ shows the condensation fields
for selected temperatures.

\section{Summing over all Matsubara frequencies}
\label{s:all}

The $\mu$-dependence of $\sigma$ and $n_{\rm B}$ found in the
preceeding section was obtained by retaining only the lowest Matsubara
frequencies.  Although this approximation is reasonable for the study
of the finite temperature phase diagram of QCD with three colors, it 
is bound to break down in the present case for small temperatures near
the condensation edge.  As $T$ becomes smaller, adjacent Matsubara
frequencies $\pm 3 \pi T$, $\pm 5 \pi T$, \ldots come closer to those
retained and eventually form a continuum as $T \to 0$. In physical
terms, the frequency continuum at $T=0$ enforces the exclusion
principle by introducing Fermi occupation factors.

As shown in the Appendix, the sum over all Matsubara frequencies leads to a
partition function of the form
\begin{eqnarray}
Z(\mu,T) = \int d\sigma \, d\Delta \,\exp 
\left\{- N \beta \Omega(\sigma,\Delta)\right\}
\label{part_all}
\end{eqnarray}
where $\beta \equiv 1/T$. The thermodynamic potential is now given by
\begin{eqnarray}
\Omega (\sigma,\Delta) & = & {\sigma^2+\Delta^2 \over  \Sigma} - 
 \sum_\pm \Big\{E_\pm(\sigma,\Delta) + 2 T 
\log[1+\exp \{-\beta E_\pm(\sigma,\Delta)\}]\Big\} + \tilde\Omega_{\rm reg} 
\, . 
\label{model:all}
\end{eqnarray} 
The excitation energies, $E_\pm(\sigma,\Delta)$, are given by
Eq.~(\ref{Epm:one}).  As before, we allow for the inclusion of a term
$\tilde\Omega_{\rm reg}$, analytic in $\mu$ and $T$, which represents
the contribution from high-energy modes. The requirement that the
$T=\mu=0$ vacuum has zero baryon density is met by taking $\tilde
\Omega_{\rm reg}$ as a constant. This constant is further set to
$\tilde\Omega_{\rm reg} = - \Sigma - 2 m$, in order for the vacuum to have
zero pressure. Again, $\Sigma$ is the chiral field for $T=\mu=m=0$.

Compared to the previous random matrix model, the full Matsubara sum has
introduced Fermi occupation factors, $f(x) = \left\{1+\exp (x/T)
\right\}^{-1}$.  This is easily seen in the two gap equations
\begin{eqnarray}
2 \, {\rm \sigma \over \Sigma} & = &  \sum_\pm {\sigma +m \pm \mu \over 
E_\pm(\sigma,\Delta)}\, \left(1 - 2 f[E_\pm(\sigma,\Delta)]\right) \\
\label{gaps_a}
2 \, {\rm \Delta \over \Sigma} & = & \sum_\pm {\Delta \over 
E_\pm(\sigma,\Delta)}\, \left(1 - 2 f[E_\pm(\sigma,\Delta)]\right).
\label{gapd_a}
\end{eqnarray}
In this form, the gap equations possess solutions which, at $T=0$ and
for all $\mu$ below the condensation edge, exhibit the same properties
as those of the vacuum, $\mu=T=0$.  For $\mu <\mu_c$, we now find a
pure chiral solution with $\Delta = 0$ and a constant chiral field
$\sigma = \Sigma$.  The baryon density vanishes exactly.  The onset
chemical potential, $\mu_c$, at $T=0$ is determined by requiring that
both gap equations are satisfied while taking $\Delta \to 0^+$.  We
find $\mu_c = (m (\Sigma +m ))^{1/2} \simeq (m \Sigma)^{1/2} +{\cal
O}(m)$.  For $\mu \ge \mu_c$, the gap equations possess a single
solution which reproduces the result of chiral perturbation theory to
lowest order in $m/\Sigma$.  Higher-order terms depend on the
functional form of Eq.~(\ref{model:all}) and are different from those
obtained in Eqs.~(\ref{mix_fields1}) and (\ref{mix_fields}). These
terms are therefore not universal.  We show the fields $\sigma$
and $\Delta$ near the condensation edge in Fig.~2$(b)$.
Figure~3$(b)$ presents their behavior as a function of $\mu$ for
selected temperatures.

The phase diagram in the full $(T,\mu)$ plane, shown in Fig.~1(b),
appears as a mere remapping of the diagram of Fig.~1(a).  Both
diagrams have the same topology. The most dramatic effect of this
remapping is a sharp vertical stretching of the condensation edges at
low and high chemical potentials.  We show in the Appendix that, at
the lower edge, the previous behavior of $\mu_c(T)-\mu_c(0)
\sim T^2$ is now replaced by essential singular behavior with $\mu_c(T) -
\mu_c(0) \sim \exp[-(\Sigma+m)/T] \sinh[\mu_c(0)/T]$. This behavior is
clearly a consequence of the introduction of Fermi occupation factors.
These factors automatically imply detailed balance, whereby two
single-quark states separated by an energy difference $\Delta E$ are
weighted by a relative factor of $\exp(-\Delta E/T)$.  Except for
these modifications at low temperature, the global picture remains
remarkably similar to that of Fig.~1(a).

\section{Building a Fermi sea}
\label{s:fermi}

Cooper pairing occurs in any attractive channel independent of its
interaction strength. Thus, even though the QCD interactions weaken on
the momentum scales appropriate for large densities, the existence of 
a diquark phase at mean field level should naturally continue to arbitrarily 
high $\mu$.  For three colors, model calculations with large $\mu$ and small 
coupling constant, $g$, find a pairing gap $\Delta \sim \mu g^{-5} 
\exp(- {\rm c}/g)$ where $c$ is a constant~\cite{Son99}.  Similar behavior 
is to be anticipated in QCD with two colors.

By contrast, the random matrix interactions saturate at large $\mu$
and do not support diquark gaps at arbitrarily high density.  This
weakening of $\left< q q \right>$ correlations is ultimately related
to the absence of a Fermi surface.  The quark states in the random
matrix model do not carry a momentum, and there is therefore no Fermi
sea.  In fact, for fixed $\sigma$ and $\Delta$, the system has only
two energy levels (the $E_\pm(\sigma, \Delta)$ in Eq.~(\ref{Epm:one})),
whereas a microscopic model exhibits a continuum of states labeled 
by their particle momenta.

In this section, we study the effects of a true Fermi sea on the
global picture established above and consider a microscopic theory
with an interaction suggested by the Nambu-Jona-Lasinio 
model~\cite{NJL}.  The structure of the interaction for large
momenta must be specified by an appropriate form factor. A sharp
three-momentum cutoff saturates the interactions at large momenta and
does not lead to a diquark gap for all $\mu$. Since we wish to explore
a model which supports pairing at all densities, we choose instead a
soft form factor,
\begin{eqnarray}
{\cal F}(q) & = & {1\over 1+(q/\Lambda)^2} \, ,
\label{form_factor}
\end{eqnarray}
which weakens the interactions with a scale of $q \sim \Lambda$.  This
choice leads to a model similar to that adopted by Berges and
Rajagopal in their study of diquark condensation for $N_c =
3$~\cite{BerRaj99}.  Our intent is to investigate general trends, and
we are not concerned with precise parameter values.  Thus, we take 
$\Lambda = 3 \Sigma$, where $\Sigma$ is again the chiral field
in vacuum for $m=0$.  The mass, $m$, is chosen as $m = 0.01 \, \Sigma$ in
order to have a clear separation between the three scales $m$,
$\sqrt{m \Sigma}$, and $\Sigma$.  (We have explored other ranges of
parameters and found the same basic trends.)  The introduction of
momentum dependence transforms the model of Eq.~(\ref{model:all}) into
\begin{eqnarray}
\Omega & = & A (\sigma^2+\Delta^2) - 
{2\over \pi^2} \sum_\pm \int_0^\infty dq\,\, q^2\,\, \Big\{
E_\pm(q,\sigma,\Delta) + 2 T \log\left(1 + 
\exp\left\{-\beta E_\pm(q,\sigma,\Delta) \right\}\right)\Big\},
\label{model:njl}
\end{eqnarray}
where the excitations energies are now momentum dependent,
\begin{eqnarray}
E_\pm(q,\sigma,\Delta) = \left((E(q)\pm \mu)^2+
{\cal F}^4(q) \Delta^2\right)^{1/2},
\label{Epm:njl}
\end{eqnarray}
with $E(q) = (q^2+(m+ \sigma {\cal F}^2(q))^2)^{1/2}$.  As before, the 
coupling constant $A$ is fixed by requiring that the vacuum chiral field be 
equal to $\sigma = \Sigma$ for $m = 0$.

The system now naturally supports a diquark condensate at all densities. To
illustrate the mechanism at work, consider first the chiral limit $m = 0$.
Then, for $T=0$ and $\mu >0$, we find that $\sigma = 0$ while $\Delta$ obeys
the gap equation
\begin{eqnarray}
A  &=& {1 \over \pi^2} \int_0^\infty dq \, q^2 \sum_\pm {\cal F}(q)^4 
{1 \over \sqrt{(q \pm \mu) ^2 + {\cal F}(q)^4 \Delta^2}}.
\end{eqnarray}
When $\Delta = 0$, the right side of this equation diverges
logarithmically due to the behavior of the integrand for $q$ near
$\mu$; it vanishes for large $\Delta$.  Hence for $A>0$, the equation
always has a solution with $\Delta > 0$. For small $\mu$, $\Delta$ is
sufficiently large that the logarithmic singularity for $q \sim \mu$ is
inoperative.  The range of contributing momenta is then determined by
the form factor ${\cal F}(q)$ and includes all $q < \Lambda$.  As $\mu$ 
increases, the momentum window in which the integrand is logarithmically 
large, $|q - \mu| \sim \Delta {\cal F}^2(\mu)$, decreases  due to the 
form factor, and the strength of the logarithmic singularity increases.  
It eventually dominates the integral at large $\mu$ and leads to 
exponentially small gaps which decrease like $\Delta(\mu) \propto 
\exp[- A /(\mu^2 {\cal  F}(\mu)^4)]$.

A similar mechanism is encountered when we depart from the chiral
limit. For $m > 0$, we have $\sigma \neq 0$, and the condensation
fields are given by the solution of two coupled gap equations.  These
equations again exhibit a logarithmic divergence as $\Delta \to 0$ 
for momenta $q$ satisfying $E(q) \simeq \mu$, where $E(q) =
(q^2+(m+ \sigma {\cal} F(q)^2)^2)^{1/2}$. We have $E(q) \ge m +
\sigma$, where the chiral field is of order $\sigma \sim
\Sigma$ for $\mu \sim \mu_c$ and decreases for larger $\mu$.  Hence,
the condition $E(q) \sim \mu$ is not obeyed for small $\mu$ as $E(q)
\sim \Sigma$ when $\mu \to 0$. It is, however, satisfied starting at
some $\mu$ in the range $\mu_c < \mu < \Sigma$ where $\sigma$ decreases 
below $\Sigma$.  For large $\mu$, the singularity condition can always 
be met.  Therefore, for $T=0$ and $\mu \ge \Sigma$, the logarithmic 
singularity is active and dominates the momentum integral.  It follows that 
$\Delta$ is always non-vanishing for moderate and large $\mu$ and that 
the diquark phase persists for all large $\mu$ as shown in Fig.~1$(c)$.

Figure~3$(c)$ shows the condensation fields for various fixed $T$.  As
we have just argued, the maximum diquark field, and therefore also the
critical temperature, decrease exponentially with $\mu$ when $\mu \ge
\Lambda$. In this region, the thermodynamic potential is dominated by
large momenta and is very sensitive to form factors.  This situtation
stands in contrast to the three color case~\cite{BerRaj99}, which has
different coupling constants and exhibits a different evolution of the
diquark field. There, most of the interesting variations in
$\Delta(\mu)$ and in the phase transition line take place below the
cutoff, $\mu < \Lambda$.  These results are likely to be much less
sensitive to form factor effects. 

Returning to the problem at hand, it is useful to remember that the present
models have been implemented at mean-field level.  Results in regions where
condensation fields are weak cannot be regarded as reliable since weak
condensates may not survive fluctuations.  An illustrative example of
the effects of long-wavelength fluctuations has recently been considered 
in lattice studies of the NJL model in two dimensions.  There, the data 
suggests a strong suppression of the diquark order parameter with
respect to the mean-field predictions and also provides indications of
strong critical fluctuations~\cite{HanLucMor00}.

Results for small $\mu$ are more reliable.  The critical physics
near the condensation edge, $\mu_c$, is primarily determined by symmetries
and is most likely protected by them.  For the NJL model, we find results
which are remarkably similar to those of the previous sections.  For zero
$T$, we find a second-order phase transition from a chiral phase to a diquark
phase at a critical chemical potential $\mu_c \sim m^{1/2}$. The chiral field
is constant for $T=0$ and $\mu < \mu_c$ and decreases roughly like $1/\mu^2$
for $\mu > \mu_c$.  The diquark field and the baryon density increase in this
region. For $T=0$ and near the condensation edge, the fields follow chiral
perturbation theory as illustrated in Fig.~2~(c). For small $T$, the critical
chemical potential behaves as $\mu_c(T) - \mu_c(0) \sim T^{3/2}
\exp\left\{-\Sigma/T\right\} \sinh[\mu_c(0)/T]$. The exponent $3/2$ in the
power law correction is directly related to the dimension of the momentum
integrals in Eq.~(\ref{model:njl}).

\section{Concluding Remarks} 
\label{s:conclusions}

The two random matrix models and the NJL model considered here all
reproduce  most the predictions of chiral perturbation theory in
the vicinity of $T=0$ and near the condensation edge, $\mu = \mu_c$.
However, the matrix models are essentially different from chiral
  perturbation theory on two respects. First, matrix models are mean
  field and do not treat the dynamics of long wavelength Goldstone excitations.
  Unlike chiral perturbation theory, the matrix models are thus not
  sensitive to effects related to spatial derivatives of the Goldstone
  fields. Second, matrix models are formulated on a large range of
  $\mu$ and $T$ which extends far beyond the condensation edge. As a
  result, they contain additional, non-universal corrections to chiral
  perturbation theory.  These corrections can, in part, be attributed
  to the effects of momentum scales larger than $m_\pi$.

This can be seen by considering both chiral limit and departures from
it.  In the strict chiral limit, $m =0$, the condensation edge goes to
$\mu_c = 0$, and much of the critical structure which we have
previously discussed disappears.  The chiral field vanishes for all
$\mu > 0$, and a finite diquark field develops for $T$ below a
critical temperature $T_c(\mu)$.  Variations in $\Delta(\mu)$ as a
function of $\mu$ occur on a large and model-dependent scale.  Away
from the chiral limit, we can distinguish two regions.  First, in the
vicinity of the condensation edge, all thermodynamic quantities follow
the predictions of chiral perturbation theory and vary with $\mu$ on a
scale $\mu_c \sim m_\pi$.  Second, far from the condensation edge, the
thermodynamic properties of all three models vary on the much larger
scales encountered in the chiral limit.  In fact, for $\mu \gg \mu_c$
the condensation fields differ from those for $m = 0$ by powers of
$m$.  The scale $m_\pi \sim m^{1/2}$, which is central to chiral
perturbation theory, thus no longer plays a role in this region where
larger and non-universal scales are dominant.

The resulting picture for $\mu \gg \mu_c$ found for each of the models 
considered here is thus quite different from that predicted by chiral 
perturbation theory.  For the random matrix models, the relevant scale of 
variation is the vacuum chiral field $\Sigma$, which is intimately related 
to the variance of the matrix elements.  This is also the scale on which 
$\left< qq \right>$ correlations weaken.  The saturation of the interactions 
manifests itself in a decrease of $\Delta(\mu)$ with $\mu$ in both random 
matrix models, compare Figs.~2~$(a)$ and $(b)$.  By contrast, the dominant 
scale of variation for $\mu \gg \mu_c$ in the NJL model is the momentum 
cutoff, $\Lambda$. This scales acts in a different way.  Because of the 
logarithmic singularity in the gap equations discussed above, $\Delta(\mu)$ 
increases until $\mu \sim \Lambda$.  For larger values of $\mu$, 
interactions weaken due to the form factors, and $\Delta$ decreases
with $\mu$.  The logarithmic singularity also causes the baryon density
to rise as $\mu$ increases from $\mu_c$.  For the NJL model, deviations 
from the results of chiral perturbation theory are opposite to those
of the random matrix models as shown in Fig.~1$(c)$.

The lesson is that large momentum scales tend to play a dominant role away 
from the condensation edge.  This should be kept in mind when interpreting
lattice data.  In this regard, it is interesting to note that the
lattice mean-field action of Dagotto et al.~\cite{DagKar86} leads to a
phase diagram with a second-order line which terminates at some high
$\mu$ and $T=0$ in a manner similar to the phase diagrams in
Figs.~1(b) and (c). This ending of the transition line is attributed
in Ref.~\cite{AloGal00} to a saturation of the lattice interactions.

A final comment concerns the evolution of the baryon density as a
function of $\mu$. We have not displayed baryon densities for $\mu \gg
\mu_c$ because random matrix models describe only the contribution of
soft modes.  Since the baryon density receives contributions from both high-
and low-energy modes, its behavior can be quite different from the
predictions of random matrix models.  Similarly, baryon densities
calculated with the NJL model are sensitive both to the form factor
chosen and to variations in the quark mass $m$ over much of the phase
diagram.  Again, this suggests that such results should be considered
with care.  By contrast, we have demonstrated that thermodynamic properties 
near the condensation edge are dominated by symmetries and that the models 
considered provide a qualitatively correct description of the baryon density 
in this region.

In conclusion, we have found that a random matrix model which retains 
only two Matsubara 
frequencies correctly reproduces the leading-order results of chiral 
perturbation theory.  Beyond leading order, however,
it produces certain unphysical results including negative baryon
densities for small $\mu$ and a chiral condensate which varies with
$\mu$.  These pathologies can be eliminated by the inclusion of all
Matsubara frequencies.  This extension of the random matrix model also
correctly produces an essential singularity in the second-order line
at $T=0$.  Neither variant of the random matrix model considered here
can support diquark condensation at arbitrarily high baryon density
since neither contains a true Fermi sea of states.  A microscopic
model capable of supporting diquark condensation at all densities does
so at the cost of introducing significant model dependence.  Near the
condensation edge, a three dimensional model also induces power law
corrections to the temperature dependence of the onset chemical
potential.  This result is robust.

The present random matrix models contain the ingredients 
  necessary for a mean-field description of the critical physics near
  the condensation edge.
All 
thermodynamic quantities in this region are dominated by those symmetries 
which we have implemented in the interactions.  Given only a minor and 
physically reasonable adjustment in the treatment of the Matsubara sum, 
the model reproduces the results of chiral perturbation theory or of any 
other microscopic model which implements the same symmetries.  Away from 
the condensation edge, however, the random matrix interactions saturate and 
lead to non-physical results.  It seems unlikely that ``improvements'' of the 
theory in this region, whether in the form of an extended random matrix 
model or of a detailed microscopic model, will be beneficial.  The large 
$\mu$ region is not dominated by symmetries, and results in that region 
will unavoidably be fragile and model dependent.  Since the random matrix 
approach is by construction free of the inevitable details of more 
microscopic models, it provides a simple and useful way of distinguishing 
between those results which are dominantly influenced by symmetries and 
those which are not.

\section*{Note added in proof}

A recent study of SU(2) lattice gauge theory with four quark flavors
by J. B.  Kogut, D. Toublan, and D. K. Sinclair~\cite{KogTouSin01} has
revealed the existence of a tricritical point at which the phase
transition changes from second to first order as temperature is
increased.  An approach based on the Landau free energy for the diquark
condensed phase shows that such point is allowed by the symmetries.
It arises from a term which contains odd powers of the order parameter
and which is related to the Goldstone fields that carry the same
quantum numbers as the diquark condensate.

The mean field level of the present random matrix models does not
include the dynamics of the Goldstone fields. Thus, our models do not
naturally lead to a tricritical point.  When interpreted as a Landau free
energy, the expression for the thermodynamical potential,
$\Omega(\sigma,\Delta)$, contains only even powers of $\Delta$ and
hence precludes the presence of such a tricritical point in the phase diagram.
 
\section*{Acknowledgements}
We thank D. Toublan, J. J. M. Verbaarschot, and A. Sch{\"a}fer for
stimulating discussions and K. Splittorff for his critical reading of
the manuscript.  We also thank the DOE Institute for Nuclear Theory at
the University of Washington for its hospitality and the Department of
Energy for partial support during the early stages of this work.

\section*{Appendix}

\subsection{Sum over all Matsubara frequencies}

Here we consider certain technical details of the random matrix model in 
which all Matsubara frequencies are included.  Following Ref.~\cite{JanNow98}, 
we introduce a Euclidean time which is the Fourier conjugate of the Matsubara 
frequency and which runs from $0 \le t \le \beta \equiv 1/T$.

Consider first the logarithmic term in $\Omega(\sigma,\Delta)$, 
Eq.~(\ref{Omega}), whose frequency trace is 
\begin{eqnarray}
{\rm Tr} \log S(\sigma,\Delta) & = & \sum_\pm \sum_{n = -\infty}^\infty
\log [\beta^2 \left((2 n + 1)^2 \pi^2 T^2 +  \Delta^2+(\sigma+m \pm
  \mu)^2\right)] \ . 
\end{eqnarray}
Here, we have included a piece of the term $\Omega_{\rm reg}$ from 
Eq.~(\ref{Omega}) by  inserting the prefactor $\beta^2 = T^{-2}$ in 
order to ensure that the argument of the logarithm is dimensionless.  
The sum over $n$ on the right side of this equation is best evaluated by 
taking the derivative of ${\rm Tr} \log S$ with respect to either 
$\sigma$ or $\Delta$.  Using then the summation formula
\begin{eqnarray}
\tanh[x] &=&  \sum_{n=-\infty}^\infty {x\over x^2 + (n+1/2)^2 \pi^2},
\end{eqnarray}
and integrating back over the condensation fields, we arrive at the 
result 
\begin{eqnarray}
{\rm Tr} \log S(\sigma,\Delta) &=& \beta \sum_\pm \left\{E_\pm(\sigma,\Delta)
  + 2 T  \log[1+\exp[-\beta E_\pm(\sigma,\Delta)]]\right\} + c(\mu,T).
\label{log_summed}
\end{eqnarray}
Here, $c$ is independent of $\sigma$ and $\Delta$ and the
$E_\pm(\sigma, \Delta)$ are the single-quark energies of
Eq.~(\ref{Epm:one}).

The frequency sum over the quadratic terms is simpler.  Since the 
condensation fields are constant in time, Fourier transformation to
Euclidean time gives
\begin{eqnarray}
{\rm Tr} A (\sigma^2 + \Delta^2 ) & = & \int_0^\beta dt 
A \left(\sigma^2+\Delta^2\right) = \beta A (\sigma^2+\Delta^2).
\label{quad_summed}
\end{eqnarray}
The saddle-point effective potential is the sum of Eqs.~(\ref{log_summed}) 
and (\ref{quad_summed}) plus an additional regular piece, $\beta \tilde 
\Omega_{\rm reg}$, which includes $c(\mu,T)$ and is chosen as follows.  
We wish to obtain a $T=\mu=0$ vacuum to have zero baryon density.  This 
constraint is met by taking $\tilde \Omega_{\rm reg}$ as a constant.  
The additional requirement that the $T=\mu=0$ vacuum should have zero
pressure sets $\tilde \Omega_{\rm reg} = - \Sigma - 2 m$, where $\Sigma$ is 
the vacuum chiral field.  Since we are primarily concerned with the 
functional form of the effective thermodynamic potential, we have chosen to
tune $A$ to $A = 1/\Sigma$ in order to obtain a chiral field $\sigma =
\Sigma$ in vacuum. This choice simplifies comparison with the other
models.   Removing a common prefactor $\beta$, we arrive at
Eqs.~(\ref{part_all}) and (\ref{model:all}).

\subsection{Temperature dependence of the onset chemical potential}

In this section, we determine the temperature dependence of the onset 
chemical potential, $\mu_c(T)$, for both the random matrix model with 
all Matsubara frequencies included and the NJL model.  We first illustrate 
the derivation for the simple random matrix model.  For fixed $T$, the 
onset chemical potential satisfies the chiral and the diquark gap equations 
with $\Delta$ set to zero.  From $\Omega (\sigma,\Delta)$ in 
Eq.~(\ref{model:all}), we have
\begin{eqnarray}
{2\over \Sigma} \,{\sigma} &  = & \tanh[{\sigma+m+\mu \over 2 T}] +
\tanh[{\sigma+m-\mu \over 2 T}],
\label{edge_chiral}\\
{2 \over \Sigma} \, \ & = &(\sigma+m+\mu)^{-1} \tanh[{\sigma+m+\mu\over 2 T}] +
(\sigma+m-\mu)^{-1} \tanh[{\sigma+m-\mu\over 2 T}].  
\label{edge_diquark}
\end{eqnarray}
The small $T$ behavior can be obtained by expanding the chiral field and the
chemical potential around their values at $T=0$ and $\mu = \mu_c$ (i.e.
$\sigma = \Sigma + \delta \sigma$ and $\mu = \mu_c +\delta \mu$) and 
expanding the explicit temperature dependence as $\tanh[x/(2 T)] 
\simeq 1- 2 \exp[- x/T]$. To first order, this expansion gives
\begin{eqnarray}
 2{\delta \sigma \over \Sigma} & \simeq & 
-4 \exp[-(\sigma+m)/T] \cosh[\mu_c /T], \\
0 & \simeq & -\delta \sigma \left\{{1\over (\Sigma+m+\mu_c)^2} + 
{1\over (\Sigma+m-\mu_c)^2} \right\} - \nonumber \\
&& \delta \mu \left\{{1 \over (\Sigma+m+\mu_c)^2} -
{1\over (\Sigma+m-\mu_c)^2} \right\} -\nonumber \\
&& 2 \exp[-(\Sigma+m)/T] \left\{{\exp[-\mu_c/T]\over \Sigma+m+\mu_c }
+{\exp[\mu_c/T]\over \Sigma+m-\mu_c} \right\}.
\end{eqnarray}
Inserting the first of these relations into the second and expanding
the denominators to first order in $\mu_c \sim m^{1/2}$ and zeroth order in
$m$, we obtain the result given in the text:
\begin{eqnarray}
\delta \mu & \simeq & \Sigma\, \exp[-{\Sigma\over T}] \sinh[{\mu_c\over T}].
\label{muTall}
\end{eqnarray}

Due to the sensitivity of the NJL model to both the mass $m$ and the cutoff
parameter $\Lambda$, it is more difficult to obtain an exact relationship
for $\mu_c(T)$ in this case.  We can, however, get a rough estimate by 
following the previous steps and restricting our attention to terms of 
zeroth order in $m/\Sigma$ and first order in $\mu_c/\Sigma$.  An expansion 
of the two gap equations to these orders then gives
\begin{eqnarray}
\delta \mu & \simeq & c \sinh[{\mu_c \over T}] \int_0^\infty dq \,q^2 
{{\cal F}(q)^4 \over
 E(q)^2} \exp[-E(q)/T] \, ,
\end{eqnarray}
where $c$ is the constant $c = (\int_0^\infty dq\, q^2 {\cal F}(q)^4 /
E(q)^3)^{-1}$ and $E(q)= (q^2+\Sigma^2 {\cal F}(q)^4)^{1/2}$ is the
single-quark energy for $\Delta = 0$ and $m = 0$. For $T \to 0$, the integral
on the right side can safely be approximated by saddle-point methods.  
Expanding $E(q)$ as $\Sigma + q^2 (1-4 \Sigma^2/\Lambda^2)/(2 \Sigma)$ and
setting $q=0$ in the denominator, we obtain
\begin{eqnarray}
\delta \mu & \simeq & c
\sqrt{\pi\over 2}\, \sinh[{\mu_c \over T}] \,\left(T\over \Sigma\right)^{3/2}\,
{\Sigma\,\exp[-{\Sigma/T}] \over (1-4 {\Sigma^2\over \Lambda^2})^{3/2}}.
\end{eqnarray}
Comparison with Eq.~(\ref{muTall}) reveals an additional power-law correction 
of $T^{3/2}$.  This correction is a consequence of having a genuine 
continuum of momentum states; the exponent $3/2$ is half the dimensionality 
of the space.

\newpage


\begin{figure}[h]

  \unitlength1.0cm
  \begin{center}
  \begin{picture}(6.0,16.0)

    \put(-2,13.2){
      \epsfysize = \grsize
      \epsfbox{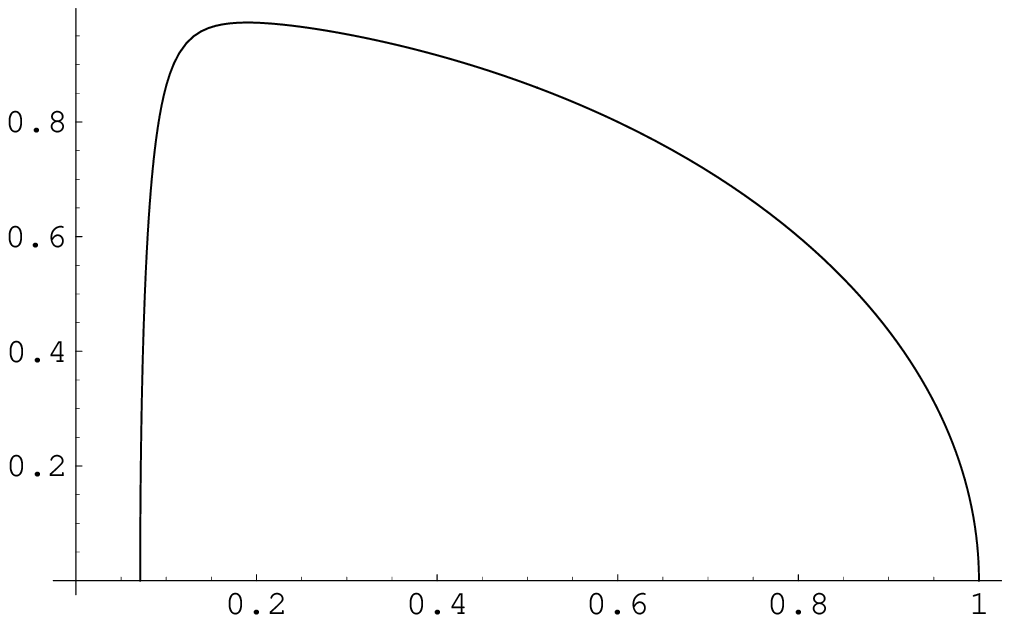}}
    \put(-2.7,16.5){$T/T_0$}
    \put(2.9,12.8){$\mu/\Sigma$}
    \put(0.,14.5){$\Delta \neq 0$}
    \put(1.,12){$(a)$}

    \put(-2,7.2){
      \epsfysize = \grsize
      \epsfbox{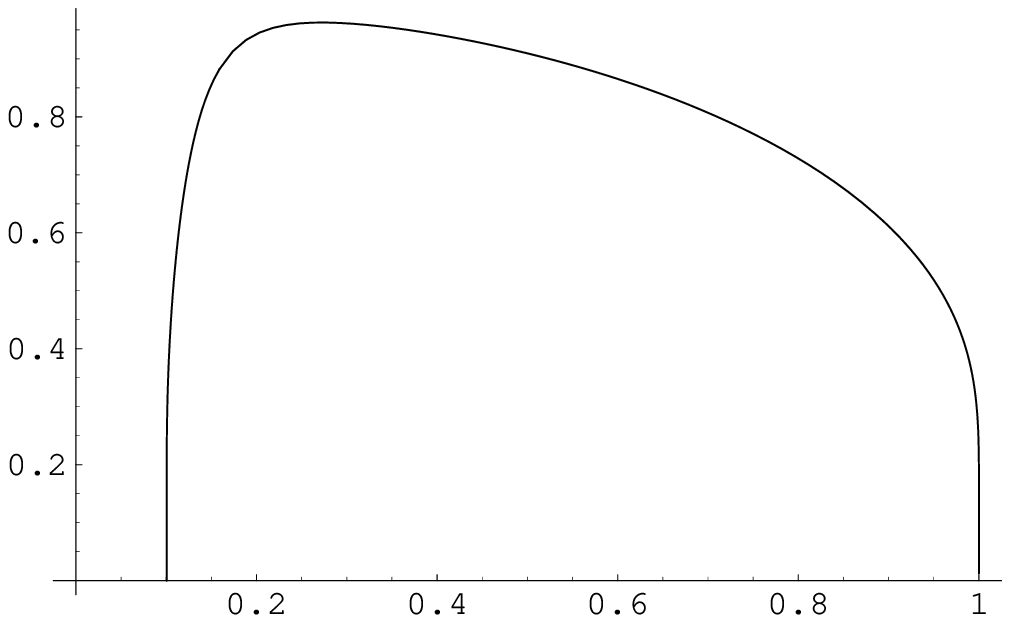}}
    \put(-2.7,10.5){$T/T_0$}
    \put(2.9,6.8){$\mu/\Sigma$}
    \put(0.,8.5){$\Delta \neq 0$}
    \put(1.0,6){$(b)$}

    \put(-2,1.2){
      \epsfysize = \grsize
      \epsfbox{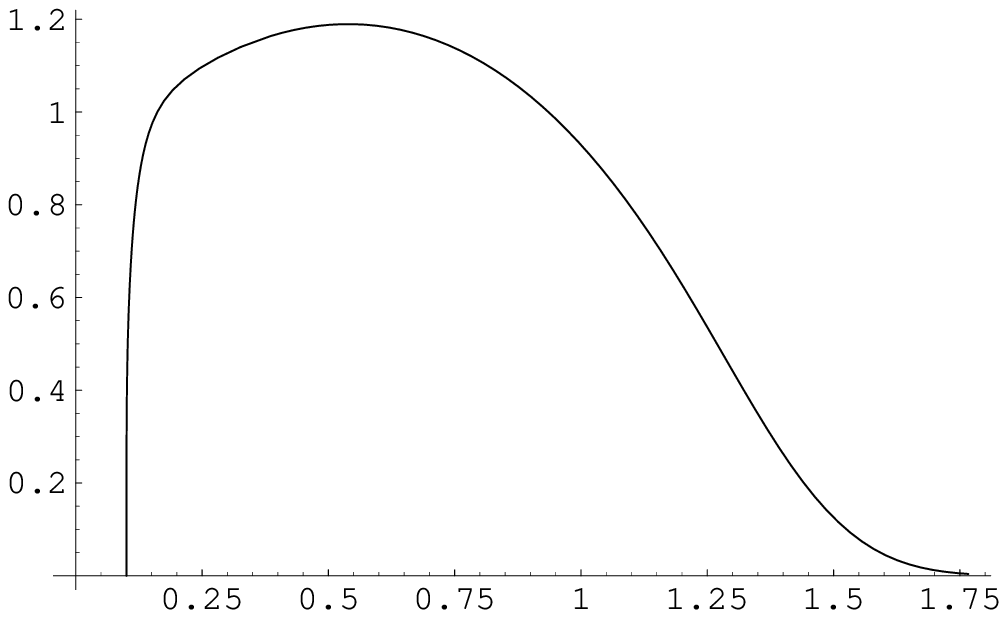}}
    \put(-2.7,4.5){$T/T_0$}
    \put(2.9,.8){$\mu/\Lambda$}
    \put(0,2.5){$\Delta\neq 0$}
    \put(1.,0){$(c)$}

  \end{picture}
  \end{center}

\caption{Phase diagrams in the $(T,\mu)$ plane for the random matrix model
  with two Matsubara frequencies included $(a)$, with all frequencies
  included $(b)$, and for the NJL model $(c)$. In each case, the
  diquark-condensed phase with $\Delta \neq 0$ is separated from that
  with $\Delta = 0$ by a second-order line. In Figs.~1 $(a)$ and
  $(b)$, $\Sigma$ is the vacuum chiral field in the limit $m=0$.
  In Fig.~1 $(c)$, $\Lambda$ is the momentum cutoff used in the NJL
  model.  In each of the three cases, $T_0$ is the critical
  temperature for $\mu = 0$ and in the chiral limit $m=0$.}

\end{figure}


\begin{figure}[h]

  \unitlength1.0cm
  \begin{center}
  \begin{picture}(6.0,16.0)

    \put(-2,13.2){
      \epsfysize = \grsize
      \epsfbox{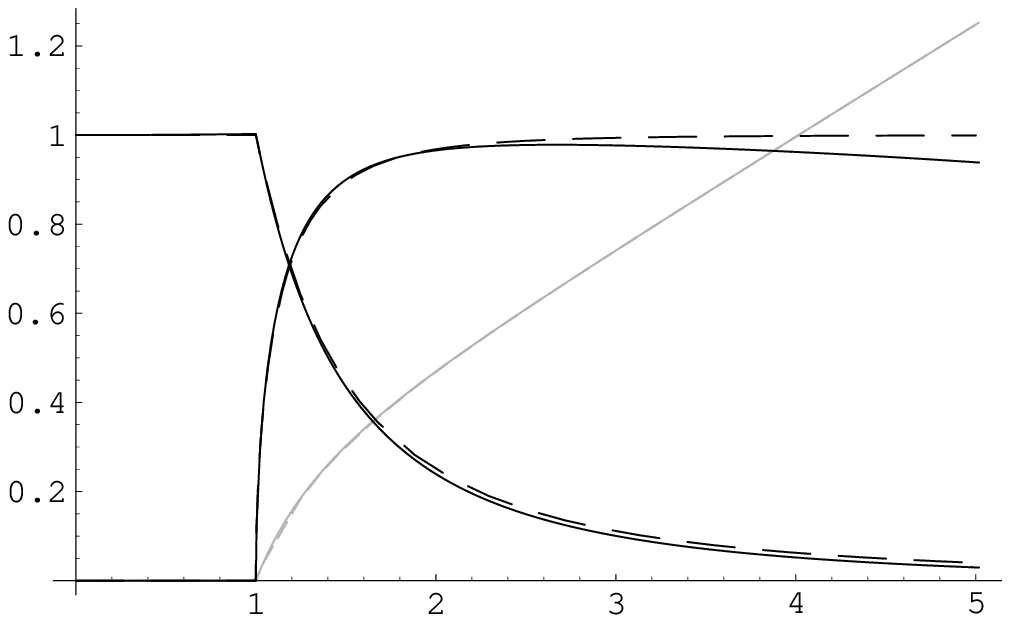}}
    \put(-1.2,16.3){$\sigma/\sigma_0$}
    \put(1.,16.3) {$\Delta/\sigma_0$}
    \put(1.3,14.8) {$n_{\rm B}$}
    \put(2.8,12.8){$\mu/\mu_c$}
    \put(1.,12){$(a)$}

    \put(-2,7.2){
      \epsfysize = \grsize
      \epsfbox{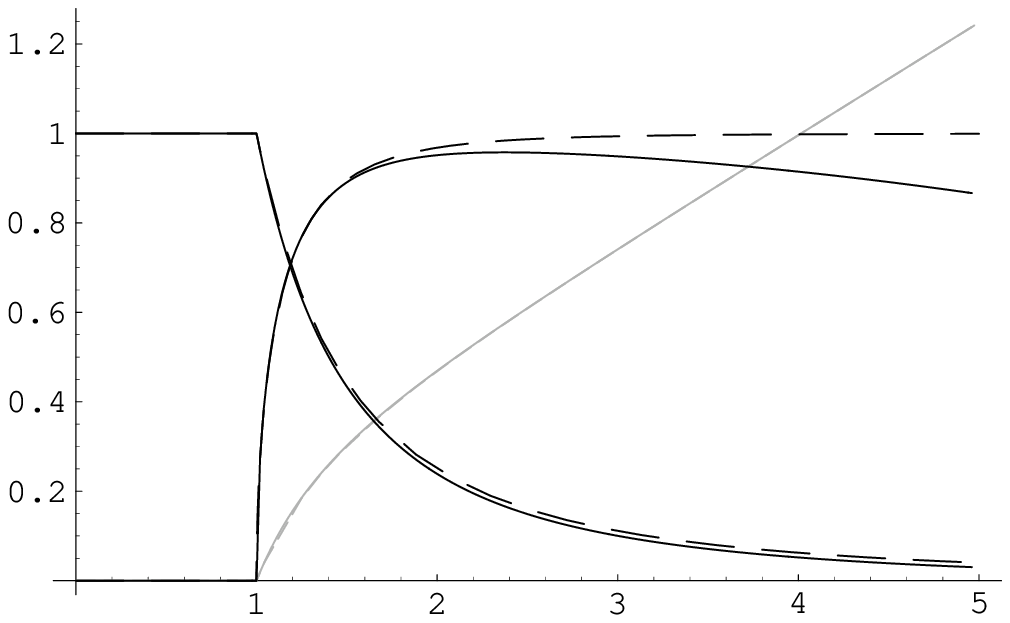}}
    \put(-1.2,10.3){$\sigma/\sigma_0$}
    \put(1.,10.3){$\Delta/\sigma_0$}
    \put(1.3,8.8){$n_{\rm B}$}
    \put(2.8,6.8){$\mu/\mu_c$}
    \put(1.0,6){$(b)$}

    \put(-2,1.2){
      \epsfysize = \grsize
      \epsfbox{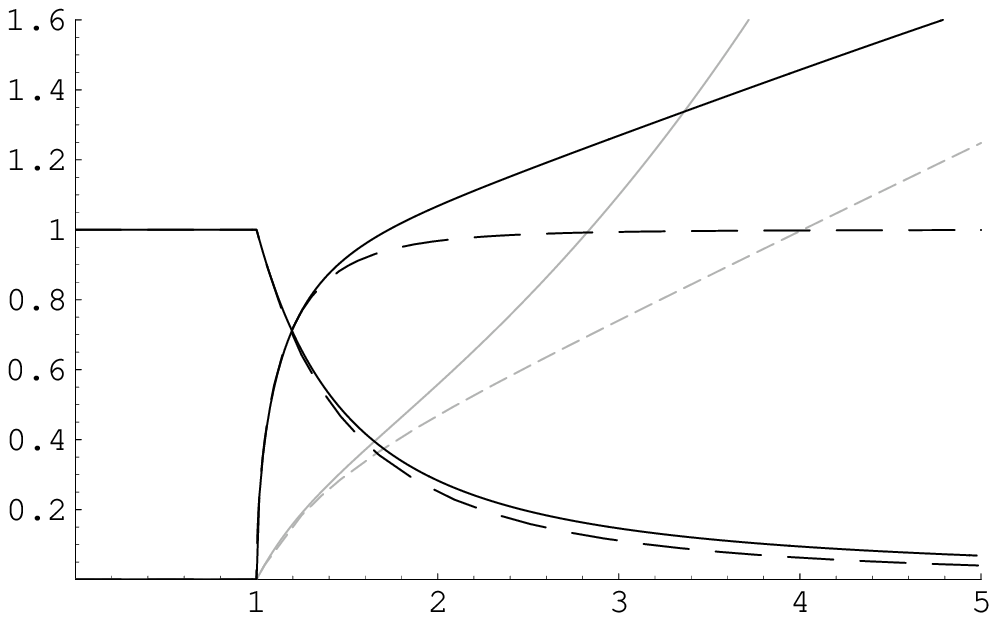}}
    \put(-1.2,3.7){$\sigma/\sigma_0$}
    \put(.7,4.1){$\Delta/\sigma_0$}
    \put(1.3,2.5){$n_{\rm B}$}
    \put(2.8,0.8){$\mu/\mu_c$}
    \put(1.,0){$(c)$}

  \end{picture}
  \end{center}

\caption{Condensation fields near the edge $(T,\mu)=(0,\mu_c)$ in the random
  matrix models with two Matsubara frequencies $(a)$, all Matsubara
  frequencies $(b)$, and in the NJL model $(c)$. In each case, the
  dashed lines represent the predictions from chiral perturbation
  theory and the continuous lines are the results from the models
  considered in this paper. In each panel, $\sigma_0$ is the chiral
  field at $\mu =0$. The grey curves represent the densities as a
  function of $\mu$, scaled so that their slope is unity at the edge
  $\mu = \mu_c$. Note that the density curves from chiral perturbation
  theory overlap with those from the random matrix models ($(a)$ and
  $(b)$).}

\end{figure}


\begin{figure}[h]

  \unitlength1.0cm
  \begin{center}
  \begin{picture}(11.0,16.0)

    \put(-2,13.2){
      \epsfysize = \grsize
      \epsfbox{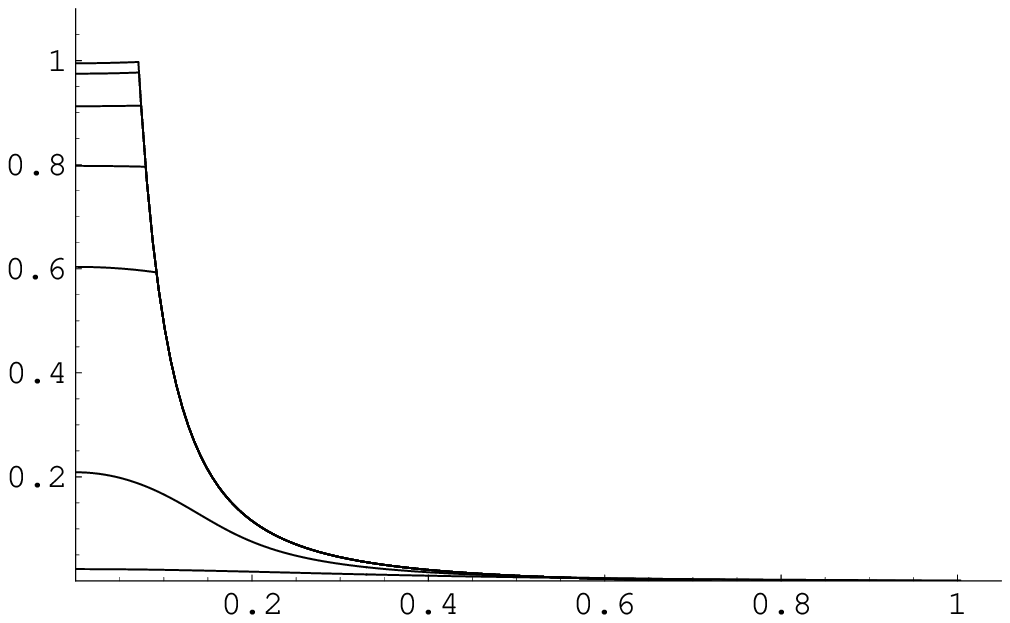}}
    \put(-2.6,16.5){$\sigma/\Sigma$}
    \put(3,12.8){$\mu/\Sigma$}
    \put(1.,12){$(a)$}

    \put(6.5,13.2){
      \epsfysize = \grsize
      \epsfbox{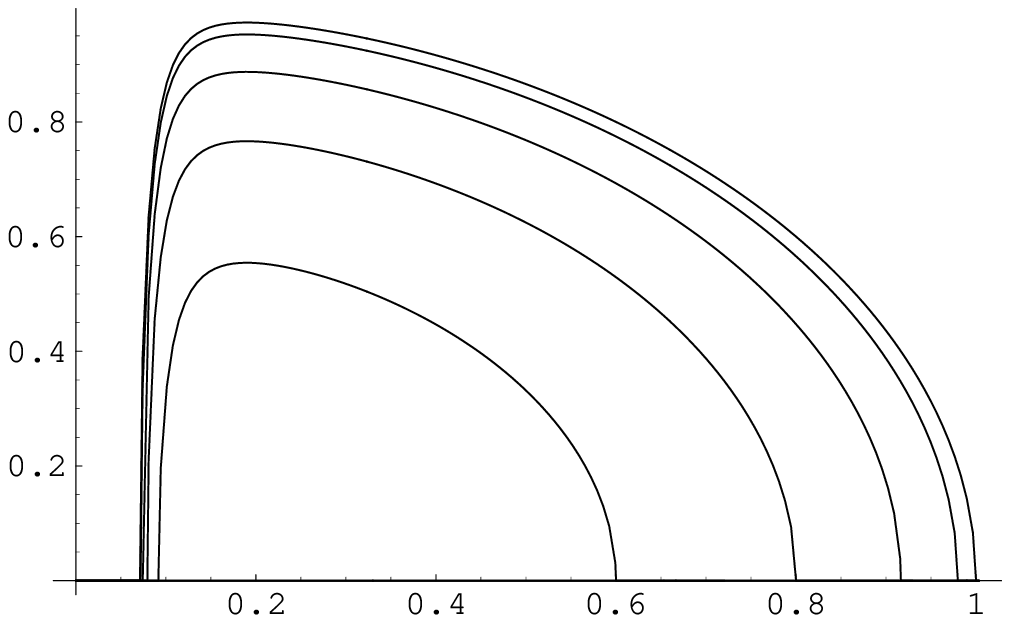}}
    \put(5.8,16.5){$\Delta/\Sigma$}
    \put(11.5,12.8){$\mu/\Sigma$}
    \put(9.5,12){$(b)$}

    \put(-2,7.2){
      \epsfysize = \grsize
      \epsfbox{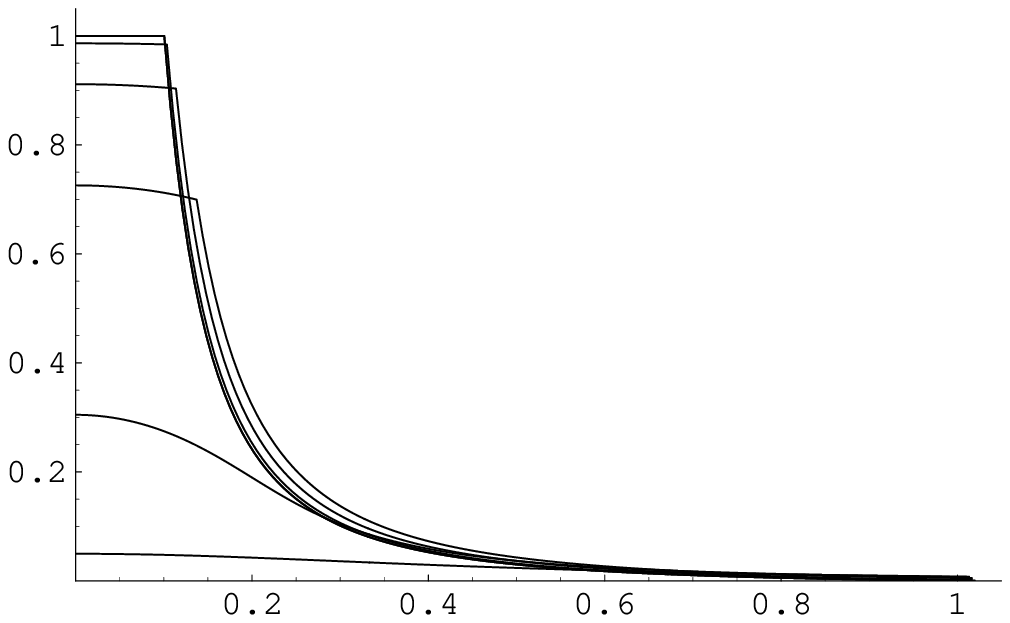}}
    \put(-2.6,10.5){$\sigma/\Sigma$}
    \put(3,6.8){$\mu/\Sigma$}
    \put(1.0,6){$(c)$}

    \put(6.5,7.2){
      \epsfysize = \grsize
      \epsfbox{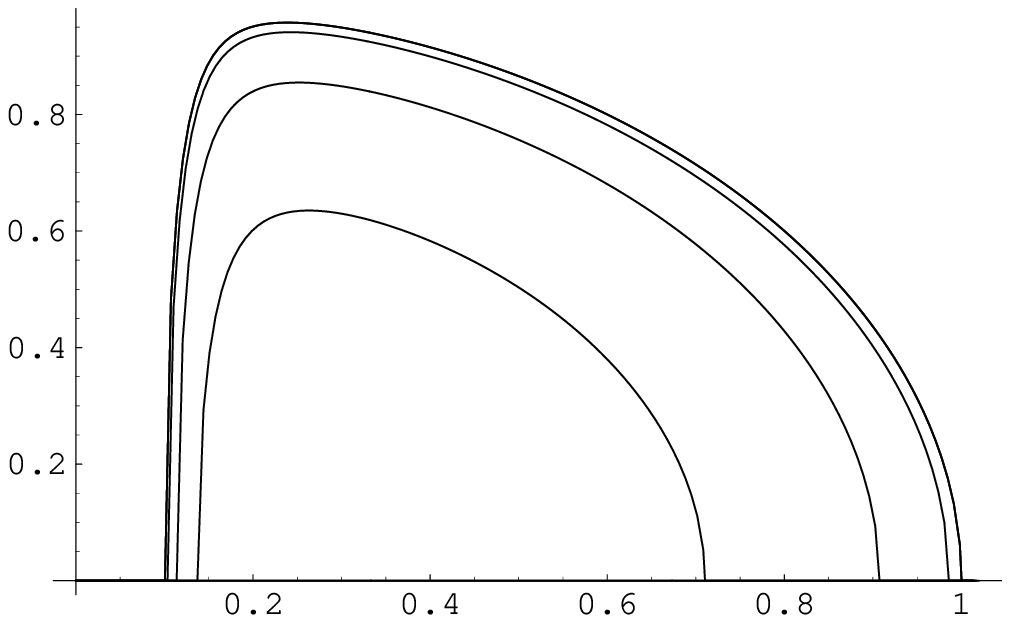}}
    \put(5.8,10.5){$\Delta/\Sigma$}
    \put(11.5,6.8){$\mu/\Sigma$}
    \put(9.5,6){$(d)$}

    \put(-2,1.2){
      \epsfysize = \grsize
      \epsfbox{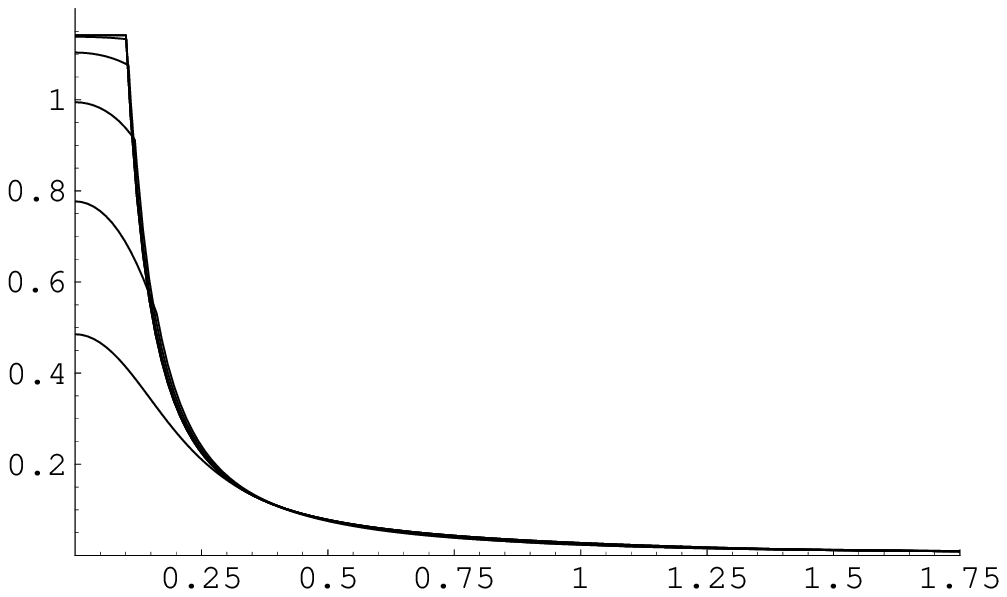}}
    \put(-2.6,4.5){$\sigma/\Sigma$}
    \put(3,.8){$\mu/\Lambda$}
    \put(1.,0){$(e)$}

    \put(6.5,1.2){
      \epsfysize = \grsize
      \epsfbox{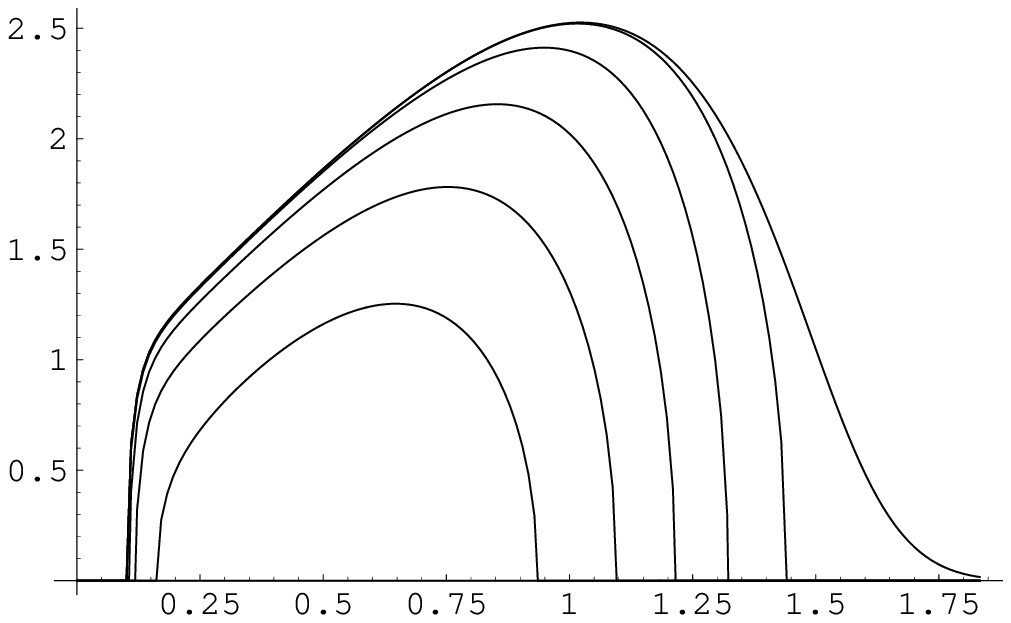}}
    \put(5.8,4.5){$\Delta/\Sigma$}
    \put(11.5,.8){$\mu/\Lambda$}
    \put(9.5,0){$(f)$}

  \end{picture}
  \end{center}

\caption{
  Chiral and diquark fields for selected temperatures in the random
  matrix model with two Matsubara frequencies included ($(a)$ and
  $(b)$), with all Matsubara frequencies included ($(c)$ and $(d)$),
  and in the NJL model ($(e)$ and $(f)$). In a given graph, the outer
  curve corresponds to $T=0$. Going inward, the inner curves
  correspond to $T/T_0 = 0.2,0.4,0.6,0.8,1.0,1.2$, where $T_0$ is the
  critical temperature for $\mu =0$ and $m=0$.  Note that the $\sigma$
  curves slightly overlap in $(c)$ for $T/T_0 = 0$ and $T/T_0=0.2$ and
  those in $(e)$ slightly overlap for $T/T_0=0,0.2$, and $T/T_0 = 0.4$.
  In Figs.~3 $(b)$, $(d)$, and $(f)$, the diquark field vanishes for
  a few of the highest selected temperatures. In each graph, $\Sigma$ is
  the vacuum chiral field in the limit $m=0$; in Figs.~3 $(e)$ and $(f)$,
  $\Lambda$ is the momentum cutoff used in the NJL model.}

\end{figure}


\begin{thebibliography}{99}
\bibitem{rem}
In the context of Ref.~\cite{CarJac99}, this means that the existence of a
pole in the current-current correlator does not depend on whether the Dyson
index $\beta$ is one (two colors), or two (three colors).
\end{thebibliography}
\end{document}